\documentstyle[11pt]{article}\textheight 230mm\textwidth 150mm
            \newcommand{\be}{\begin{eqnarray}}
            \newcommand{\ee}{\end{eqnarray}}
\newcommand{\e}[1]{\label{e:#1}\end{eqnarray}}
     \newcommand{\eg}{{\em e.g.\ }}
            \newcommand{\ie}{{\em i.e.\ }}
            \newcommand{\ga}{{\gamma}}
 \newcommand{\Ga}{{\Gamma}}
            \newcommand{\la}{{\lambda}}
            
            \newcommand{\del}{{\delta}}
           \newcommand{\ra}{{\rightarrow}}
 \newcommand{\lea}{{\leftarrow}}
            \newcommand{\lra}{{\leftrightarrow}}
            \newcommand{\Lra}{{\Leftrightarrow}}

            \newcommand{\beq}{\begin{quote}}
            \newcommand{\eq}{\end{quote}}
            
            \newcommand{\al}{\alpha}
            \newcommand{\half}{\frac{1}{2}}
            \newcommand{\ben}{\begin{enumerate}}
            \newcommand{\een}{\end{enumerate}}
\newcommand{\bea}{\begin{array}}
            \newcommand{\ea}{\end{array}}
            \newcommand{\bit}{\begin{itemize}}
            \newcommand{\ei}{\end{itemize}}
\newcommand{\ve}{{\varepsilon}}
    		
            \newcommand{\nn}{\nonumber}
   
            \newcommand{\r}[1]{(\ref{e:#1})}
            \newcommand{\edfl}[1]{\label{#1}\end{df}}
\def\L{\left}
\def\R{\right}
\def\tilde{\widetilde}
\def\bar{\overline}
\def\d{{\partial}}
\def\JMP{{\sl J.\ Math.\ Phys.}}

\def\cF{{\cal F}}

\def\cH{{\cal H}}

\def\cL{{\cal L}}
\def\cM{{\cal M}}

\def\cU{{\cal U}}
\def\cV{{\hat{V}}}
\def\caV{{\cal V}}
\def\cW{{\cal W}}
\def\cX{{\cal X}}
\def\tX{{\tilde{X}}}

\def\tGamma{{\tilde{\Gamma}}}

\def\IJMPA{{\sl Int. J. Mod. Phys. }A}
\def\MPLA{{\sl Mod. Phys. Lett.} A}

\begin{document}
            \begin{titlepage}
            \newpage
            \noindent
            G\"oteborg ITP 95-28\\
            October 1995\\
												hep-th/9510201
            \vspace*{35 mm}
            \begin{center}{\LARGE\bf General triplectic quantization}
\end{center}
            \vspace*{15 mm}
        \begin{center}{\large \bf Igor Batalin}
\footnote{On leave of absence from
P.N.Lebedev Physical Institute, 117924  Moscow, Russia\\ .\hspace{3.5mm}
 E-mail:
batalin@lpi.ac.ru } and {\large \bf Robert Marnelius}
\footnote{E-mail:
tferm@fy.chalmers.se} \\
          \vspace*{10 mm} {\sl
            Institute of Theoretical Physics\\ Chalmers University of
            Technology\\ S-412 96  G\"{o}teborg, Sweden}\end{center}
            \vspace*{27 mm}
            \begin{abstract}
            The general structure of the $Sp(2)$ covariant version
of the
field-antifield  quantization of general constrained systems in the
Lagrangian formalism, the so called triplectic quantization,
as presented in our previous
paper with A.M.Semikhatov is further generalized and clarified.
 We present new unified
expressions for the generating operators which are more invariant
and which
yield  a natural realization of the operator $V^a$ and provide for a
geometrical explanation for its presence. This $V^a$ operator provides
then for an
invariant definition of a degenerate Poisson bracket on the triplectic
 manifold being
nondegenerate on  a naturally defined submanifold. We also define
inverses to
nondegenerate antitriplectic metrics and  give a natural
generalization of the
conventional calculus of exterior differential forms which \eg
explains the properties
of these inverses. Finally we define and give a consistent
treatment of second class
hyperconstraints.  \end{abstract} \end{titlepage}
            \newpage
            \setcounter{page}{1}
            \section{Introduction}
A general $Sp(2)$-covariant Lagrangian quantization of general
gauge theories was
presented in \cite{Trip} and was  nick-named "triplectic
quantization". The paper was
based on the  $Sp(2)$-symmetric  formalism  developed  in
\cite{BLT-2}
--\cite{BM} which had its origin
in a BRST-antiBRST Hamiltonian formalism. (For other works on the
$Sp(2)$ formalism
see also \cite{Hull}
--\cite{OMK}.) In \cite{Trip} almost all general properties
of the antisymplectic
formalism was generalized to the $Sp(2)$-case. In fact, even
the antisymplectic
formalism was slightly generalized.

The  purpose of the present paper is to make some improvements,
clarifications and
further generalizations of the triplectic quantization
as presented in  \cite{Trip}.
Our main new results are new unified expressions for the
generating operators
$\Delta^a_\pm$, expressions which are more invariant and
which provide for a natural
covariant realization of the $V^a$ operator and a geometrical
explanation for its
presence in the $Sp(2)$ covariant formulation. We also show that
the allowed form of
$V^a$ is more general than those considered in \cite{Trip}. In fact,
the natural
arbitrariness of $V^a$ is parametrized by a bosonic function which
is arbitrary up to
the requirement that it must satisfy the classical master equation.
Different choices
of $V^a$ correspond then to different boundary conditions on the
master action. Thus, we
resolve  the clash between the $V^a$ in \cite{BLT-2}-\cite{BLT-4}
and in
\cite{Trip},\cite{BM}. The interesting Poisson bracket introduced
in \cite{Trip}, which
is defined on a submanifold of the triplectic manifold, is
generalized to include more
general $V^a$ operators. In fact, the unified expressions
for $\Delta^a_\pm$ allow us
to give a completely invariant definition of a degenerate
Poisson bracket
 on the triplectic manifold which is
nondegenerate on  a naturally defined submanifold.

The nondegeneracy properties of the antitriplectic metric
are here specified by the
requirement  of the existence of an inverse which at first
glance look rather
peculiar. However, by means of a generalized exterior
differential form calculus,
which we introduce, this inverse may be written as coefficients
of a two-form in
terms of which its properties become natural ones. The
exterior differential is
here an $Sp(2)$ vector which satisfies the same algebraic
properties as the
generating operators $\Delta_\pm^a$. Our specification of an
inverse to a nondegenerate
antitriplectic metric allows us also to introduce  second
class hyperconstraints and
to define the triplectic counterpart of the Dirac bracket.

The paper is organized as follows:
In section 2 we give  the general properties of the basic
objects in triplectic
quantization generalizing the results of \cite{Trip} to
general $V^a$ operators.
Inverses to triplectic metric and generalized differential
forms are introduced.
In section 3 we present our new unified expressions for the
generating operators
which are more invariant and which lead to natural realizations
of $V^a$ and explain
their origin. In section 4 we point out how the gauge fixing
procedure of \cite{Trip} is
generalized to more general $V^a$ operators, and finally in
section 5  we present the
treatment of triplectic second class constraints. In an
appendix we give some basic
objects in terms of Darboux coordinates.

\setcounter{equation}{0}
\section{Basics of general triplectic quantization}

In general triplectic quantization one considers a
triplectic  manifold  $\cM$
with local coordinates $\Gamma^A$,
$A=1,\ldots,6N$, with Grassmann parities
$\ve(\Gamma^A)\equiv\ve_A\in\{0,1\}$.
$\cM$ is endowed with a volume measure
$d\mu(\Gamma)\equiv\rho(\Gamma)[d\Gamma]$ where
$\rho(\Gamma)$ is a scalar density. On $\cM$ we have
the basic   tensor
$E^{ABa}(\Gamma)$,  $a=1,2$, the antitriplectic metric, satisfying
the properties
\be
&&E^{A B a}={}-E^{B A a}(-1)^{(\ve_A+1)(\ve_B+1)},\;\;\;
\ve(E^{A B a})=\ve_A+\ve_B+1\,.
\e{201}
It is
required to be nondegenerate in the sense that there
should exist a tensor $Y^a_{bcBC}$
with the properties
\be
&&\ve(Y^a_{bcBC})=\ve_B+\ve_C+1,\;\;\;Y^a_{bcBC}=
-Y^a_{cbCB}(-1)^{\ve_B\ve_C}\,,
\e{202}
satisfying
\be
&&E^{ABb}Y^a_{bcBC}=\del^A_C\del^a_c\,.
\e{203}
 It follows then  from \r{201}, \r{202} and
\r{203} that the conjugate
equation  holds as well, \ie
\be
&&Y^a_{cbCB}E^{BAb}=\del^A_C\del^a_c\,.
\e{204}
Notice, however,  that $Y^a_{bcBC}$ is not uniquely
determined by these conditions.

In triplectic quantization there are two basic pairs of odd
(fermionic)  differential
operators, $\Delta_+^a$ and $\Delta_-^a$,
which generate the master equations.
Their dependence on $\hbar$ is given by
 \be
&&\Delta_\pm^a\equiv\Delta^a\pm{i\over\hbar}\cV^a,\;\;\;a=1,2 \,,
\e{205}
where $\Delta^a$ are second order differential
operators with respect to $\Gamma^A$,
and $\cV^a$ first order ones. $\Delta^a$ and $\cV^a$
are defined by the expressions
\be
&\Delta^a=\half(-1)^{\ve_A}\rho^{-1}\d_A\!\circ\!
\rho E^{A B a}\d_B,\;\;\;a=1,2\,,
\e{206}
\be
&\cV^a\equiv V^a+\half{\rm div}V^a,\;\;\;a=1,2\,,\nn\\
&V^a\equiv(-1)^{\ve_A}V^{Aa}\d_A ,\;\;\;
\ve(V^{Aa})=\ve_A+1\,,\nn\\
&{\rm div}V^a\equiv\rho^{-1}\L(\d_A\rho
V^{Aa}\R)(-1)^{\ve_A}\,.
\e{207}
The factor $(-1)^{\ve_A}$ in the expression for $V^a$
is inserted for convenience.
Notice also that the antisymmetry properties \r{201} of $E^{ABa}$
 just provide for the required symmetry  of  the coefficients for
$\d_B\d_A$ in  \r{206}.

$\Delta_\pm^a$ are hermitian operators with respect
to the inner product
\be
&&(f,g)\equiv\int f^*(\Gamma)g(\Gamma)d\mu(\Gamma)
\e{2071}
for real coordinates, or provided one imposes the
complex structure induced by the
differential complex-conjugate conditions
 \be
&(d\Gamma^A)^*=d\Gamma^BI^A_{\;B},\;\;\;\d_B=I^A_{\;B}\d_A^*\,,
\e{2072}
where $I^A_{\;B}$ are fields which satisfy the conditions
\be
&(I^A_{\;B})^*I^B_{\;C}(-1)^{\ve_C(1+\ve_B)}=
\del^A_{\;C}\;\Rightarrow\;|{\rm sdet \ }
I|^2=1\,,
\e{2073}
\be
&\d_CI^A_{\;B}-\d_BI^A_{\;C}(-1)^{\ve_B\ve_C}=0\,,
\e{2074}
\be
&\d^*_BI^B_{\;A}(-1)^{\ve_B(\ve_A+1)}=\d_A\ln {\rm sdet \ } I=0\,.
\e{2075}
The last relation follows also from the requirement
of a real measure (${\rm sdet \ }
I=1$). Under these conditions  we have
\be
&(E^{CDa})^*=(-1)^{\ve_A(\ve_D+1)}I^D_{\;A}E^{ABa}I^C_{\;B}
,\;\;\;(V^{Aa})^*=(-1)^{\ve_B}V^{Ba}I^A_{\;B}\,.
\e{2076}
 Although we shall not make use of this complex structure
in the following it should be useful to know that it
may always be imposed.

The basic odd operators \r{205} are required to satisfy
\be
&&
[\Delta_\pm^a,\Delta_\pm^b]=0\quad\Leftrightarrow\quad
\Delta_\pm^{\{a}\Delta_\pm^{b\}}=0\,,
\e{208}
where  the commutator $[\;,\;]$ here and in the sequel
denotes the graded supercommutator \
$[A,B]=AB-(-1)^{\ve(A)\ve(B)}BA$, and where the curly
bracket indicates
symmetrization in the indices $a$ and $b$.
These conditions imply in turn the following
conditions on $\Delta^a$ and $\cV^a$ by
identifications of different powers of $i/\hbar$ in
\r{208} (The conditions from
$\Delta_+^a$ and $\Delta_-^a$ are identical.) \be
&&[\Delta^a,\Delta^b]=0\quad\Leftrightarrow\quad
\Delta^{\{a}\Delta^{b\}}=0\,,
\e{209}
\be
&&[\Delta^{\{a},\cV^{b\}}]=0\;\Lra\;\Delta^{\{a}\cV^{b\}}+
\cV^{\{a}\Delta^{b\}}=0\,,
\e{211}
and
 \be
&&[\cV^a,\cV^b]=0\quad\Leftrightarrow\quad\cV^{\{a}\cV^{b\}}=0\,.
\e{210}
Identifying furthermore the
coefficients for different powers of $\d_A$ we find
that the conditions \r{209}
imply  \be
&&
E^{A D\{a} \d_D E^{B C|b\}}(-1)^{(\ve_A+1)(\ve_C+1)} + {\rm
cycle}(A\,B\,C)=0\,,
\e{212}
\be
&&(-1)^{\ve_A}\rho^{-1}\d_A\Bigl(\rho E^{A B \{a}\d_B\,
(-1)^{\ve_C}\rho^{-1}\d_C\bigl(\rho E^{C D|b\}}\bigr)\Bigr)=0\,.
\e{213}
They are conditions on $E^{ABa}$ and $\rho$. One may
 notice that \r{212} imply the
following conditions on $Y^{a}_{bcBC}$ in \r{203}:
\be
&&(-1)^{\ve_C\ve_A}E^{FBb}E^{GAa}
\Bigl(\del_a^{\{f}\d_AY^{g\}}_{bcBC}(-1)^{\ve_C\ve_A}
+\nn\\
&&\Bigl.+{\rm
cycle}[(a,A), (b,B), (c,C)]\Bigr)E^{CHc}(-1)^{(\ve_G+1)
(\ve_H+\ve_F+\ve_B)}=0\,.
\e{2131}
These conditions may easily be understood in terms of
appropriately  generalized
differential forms. The basic objects for such a
form calculus are the exterior
differentials $d^a$, $a=1,2$,   satisfying
\be
&&[d^a, d^b]=0\;\Leftrightarrow\;d^{\{a}d^{b\}}=0\,.
\e{2132}
Notice the algebraic similarity with the properties \r{208}  of
$\Delta^a_{\pm}$.
In terms of these differentials we may  define
differential forms which are also
$Sp(2)$ tensors, and \eg a vector form $\omega^a$ is
 closed if $d^{\{a}\omega^{b\}}=0$
and exact if $\omega^a=d^{a}\sigma$ where $\sigma$
is an $Sp(2)$ scalar form of one
lower degree.  Now $Y^{a}_{bcBC}$ may be expressed
as coefficients of  such a
two-form (our conventions correspond to those of 
ref.\cite{Ber}): \be
&&\omega^a_2=\half Y^{a}_{bcBC}(-1)^{\varepsilon_C}
\,d^c\Gamma^C\wedge d^b\Gamma^B-F_{bB}D^ad^b\Gamma^B\,.
\e{2133}
where $F_{aA}$ is an arbitrary vector field with 
Grassmann parity $\ve(F_{aA})=\ve_A+1$. $D^a$ is a 
covariant exterior differential.
This explains the existence of a $Y^{a}_{bcBC}$
 with the properties \r{202}.
 It follows  that the factors within the parenthesis
 of \r{2131}
are  equal to  the coefficients of $d^{\{f}
\omega_2^{g\}}$. However,
since we are unable to remove the tensors $E^{FBb}$,
 $E^{GAa}$ and $E^{CHc}$
eq.\r{2131} does not imply that $\omega^a_2$ is closed.
This is related to the fact
that $Y^{a}_{bcBC}$ is not unique. The last 
term in \r{2133} is necessary in order
to make  $\omega^a_2$  exact when \be
&&Y^{a}_{bcBC}=\del^a_b D_BF_{cC}-
\del^a_c D_CF_{bB}(-1)^{\ve_B\ve_C}\,,
\e{2134}
where $D_A$ is a covariant derivative with symmetric 
connections determined by
\be
&&D_AE^{BC a}=0,\;\;\;[D_A, D_B]=0
\e{21341}
 The expression \r{2134} is 
a particular solution of \r{2131} for which we have
\be
&&\omega^a_2=d^a\omega,\;\;\;\omega=
F_{aA}d^a\Gamma^A(-1)^{\ve_A}\,.
\e{2135}

In terms of $V^a$  in \r{207} , and by identifying the
coefficients for different powers of $\d_A$ we find
that conditions \r{211} imply
\be
(\Delta^{\{a}{\rm div}V^{b\}})=0,
\e{215}
\be
\L(\Delta^{\{a}V^{B|b\}}
(-1)^{\ve_B}\R)+\half\L(V^{\{a}(-1)^{\ve_A}
\L(\rho^{-1}\d_A\rho
E^{AB|b\}}\R)\R)+\half E^{BA\{a}(\d_A{\rm div}V^{b\}})=0,
\e{216}
\be
(V^{\{a}E^{AB|b\}})=(-1)^{\ve_A}V^{A\{a}
\stackrel{\lea}{\partial}_CE^{CB|b\}}-(A\lra
B)(-1)^{(\ve_A+1)(\ve_B+1)},
\e{217}
and that \r{210} implies
\be
&&[V^{\{a},
{\rm div}V^{b\}}]=0\;\;\Leftrightarrow\;
\;(V^{\{a}{\rm div}V^{b\}})=0,
 \e{214}
\be
&&[V^a,
V^b]=0\;\;\Leftrightarrow\;\;(V^{\{a}V^{Bb\}})=0.
\e{2141}
Only the last conditions are conditions on $V^a$
itself. Eq.\r{217} involves also
$E^{ABa}$ and \r{215},\r{216} and \r{214} are
conditions on $\rho$, $E^{ABa}$ and
$V^a$.

In triplectic quantization there are two
 antibrackets, $(\;,\;)^a$, $a=1,2$, and they are
defined through the relation
\be
&&\Delta^a(F\,G)=(\Delta^aF)\,G +
 F\,\Delta^aG\,(-1)^{\ve(F)} +
(F,G)^a(-1)^{\ve(F)}\,,
\e{221}
where $F$ and $G$ are functions on $\cM$.
This combined with \r{206} leads to the
following expression for the antibrackets in \r{221}
\be
&&(F,G)^a={F}{{\stackrel{\lea}{\partial_A}}}
E^{A B a}{{\partial_B}} {G} ,\;\;\;a=1,2\,.
\e{222}
{}From the properties \r{201}, \r{212} and
\r{213} of the tensor $E^{ABa}$ it
follows that these antibrackets satisfy the properties
\be
&&(F,G)^a = - (G,F)^a
(-1)^{(\ve(F)+1)(\ve(G)+1)}\,,
\e{223}
\be
&&\ve((F,G)^a)=\ve(F)+\ve(G)+1\,,
\e{224}
\be
&&(F,GH)^a=(F,G)^aH+G(F,H)^a(-1)^{(\ve(F)+1)\ve(G)}\,.
\e{225}
The identities $\Delta^{\{a}\Delta^{b\}}(FG)\equiv
0$ and $\Delta^{\{a}\Delta^{b\}}(FGH)\equiv
0$ together with \r{212}, \r{213}, and \r{222}
 yield furthermore
\be
&&\Delta^{\{a}(F,G)^{b\}}=(\Delta^{\{a}F,G)^{b\}} +
 (F,\Delta^{\{a}G)^{b\}}(-1)^{\ve(F)+1}
\e{226}
and
\be
((F,G)^{\{a},H)^{b\}}(-1)^{(\ve(F)+1)(\ve(H)+1)} +
{\rm cycle}(F,G,H) =0
\e{227}
respectively.  Equation \r{227} is a version of
the Jacobi identities satisfied by the
two antibrackets.
 Notice also that the equality
\r{217} is equivalent to the property \be
&&V^{\{a}(F, G)^{b\}}=(V^{\{a}F, G)^{b\}}+(F,
 V^{\{a}G)^{b\}}(-1)^{\ve(F)+1}\,.
\e{229}
Thus, the $V^a$ used here belongs to a more
general class than the one of \cite{Trip}
where $V^a$ was chosen to satisfy this relation
 without symmetrization in $a$ and $b$.
In fact, in \cite{Trip} $V^a$ were required to
satisfy the relation
\be
&&\epsilon_{ab}[\Delta_+^a, \Delta_-^b]=0\;
\Leftrightarrow\;\epsilon_{ab}[\Delta^a,
\cV^b]=0\,. \e{2291}
Although this was a convenient restriction
it is not a necessary one. We shall refer
to the $V^a$ of \cite{Trip} satisfying
\r{2291} as the symmetric $V^a$. (In \cite{ND} 
another argument for the general
relation \r{229} was given.)

The above conditions do not determine
a unique  $V^a$.  In fact, its
natural arbitrariness is given by
transformations of the form
\be
&&V^a\mapsto V'^a=V^a+(\cH, \cdot)^a
\e{238}
where $\cH(\Gamma)$ is a bosonic function
which is a solution of the
 equation
\be
&&\half(\cH, \cH)^a+V^a\cH=0\,.
\e{239}
This restriction  on $\cH$ follows
 from the  conditions
\be
&&[\cV'^a, \cV'^b]=0,
\e{241}
while
\be
&&[\Delta^{\{a}, \cV'^{b\}}]=0
\e{2401}
allows for arbitrary $\cH$.
The transformation \r{238} also induces
\be
&&\half{\rm div}\,V'^a=\half\,
{\rm div}\,V^a+\Delta^a\cH\,.
\e{240}

The basic starting point in general
triplectic quantization is the
determination of the quantum master
action $W(\Gamma;\hbar)$ where
$W(\Gamma;\hbar)$ is expandable in
powers of $\hbar$ and is required to satisfy the
quantum master equation
\be
&&\Delta_+^a\exp\L\{{i\over\hbar}W\R\}=0\,,
\e{2411}
which may be equivalently written as
\be
\half(W,\,W)^a + V^aW =i\hbar\Delta^aW+\half
i\hbar\,{\rm div}\,V^a\,.
 \e{2412}
Notice that \r{208} are then just
compatibility conditions for \r{2411}.
The master action $W$ should also be
specified by imposing boundary conditions
requiring them to coincide at $\hbar=0$
 with the original action $S$ of the theory on
some Lagrangian submanifold $\cL_0$:
\be
&&\L.W(\cdot;0)\R|_{\cL_0}=S(\cdot)\,.
\e{2413}

{}From \r{2412} it follows  that $\cH(\Gamma)$
in the transformation \r{238} satisfies
the classical master equation. This in turn
implies that the quantum master equation
\r{2411} may also  be written as  \be
&&{\Delta'}_+^a\exp{\L\{{i\over\hbar}W'\R\}}=0\,,
\e{242}
where
\be
&&{\Delta'}_+^a={\Delta}^a+{i\over\hbar}
\cV'^a,\;\;\;W'=W-\cH\,.
\e{243}
Hence, the quantum master equation \r{2411}
is invariant under transformations
of the form \r{238} provided the action $W$ at the
same time is shifted to $W'$ given by \r{243}.
In particular, $V^a$  may be transformed away
completely if it is Hamiltonian with
respect to the antibracket. In order to exclude
this possibility we need to define
the nondegeneracy properties of $V^a$.

In \cite{Trip} an interesting Poisson bracket
 was introduced on a submanifold of
$\cM$ in terms of which the nondegeneracy
 conditions on symmetric $V^a$ operators was
implicitly defined by the requirement that
the Poisson bracket is nondegenerate on
this submanifold. Such a Poisson bracket may
 also be introduced for more general $V^a$
operators not satisfying \r{2291}. Consider
 the bracket
\be
&&\{F,G\}\equiv\half
\epsilon_{ab}\L( (F, V^bG)^a+
(-1)^{\ve(F)}(V^bF, G)^a\R)\,,
\e{230}
 which is a
generalization of the corresponding
definition in \cite{Trip}.  This bracket does not
satisfy a Lie superalgebra  on the whole
triplectic manifold $\cM$.
We have to consider a submanifold of $\cM$.
 We require therefore $V^a$ to be such that
there exists a class of functions $\cF$
satisfying
\be
&&(F, G)^a=0,\;\;\;\forall F,G\in\cF\,,
\e{231}
\be
&&(F, V^{\{a}G)^{b\}}+(-1)^{\ve(F)}
(V^{\{a}F, G)^{b\}}=0,\;\;\;\forall F,G\in\cF\,,
\e{232}
and also
\be
&& \epsilon_{ab}\L((F, (\cH,G)^{b})^{a}+
(-1)^{\ve(F)}((\cH,F)^{b}, G)^{a}\R)=0
,\;\;\;\forall F,G\in\cF\,,
\e{233}
\be
&&(F, (\cH, G)^{\{a})^{b\}}+
(-1)^{\ve(F)}((\cH,F)^{\{a}, G)^{b\}}=0,\;\;\;\forall
F,G\in\cF\,, \e{234}
for bosonic functions $\cH$ satisfying
the classical master equation \r{239}.
The definition \r{230} and the conditions
\r{232} are then invariant under the
transformations
\be
&&V^a\mapsto{V'}^a=V^a+(\cH,\cdot)^a
\e{2341}
for those $\cH$ which satisfy
\r{233},\r{234}. If also
\be
&&\{F, G\}\in\cF,\;\;\;\forall F,G\in\cF\,,
\e{2342}
and if we assume that there is a symmetric
$V^a$ in the class of $V^a$ operators
determine by the equivalence relation
\r{2341} then the bracket
\r{230} is a graded Poisson bracket on $\cF$,
 \ie it satisfies the properties
\be
&&\{G,F\}=-(-1)^{\ve(F)\ve(G)}\{F,G\}\,,
\e{235}
\be
&&\{F, GH\}=\{F, G\}H+G\{F, H\}(-1)^{\ve(F)\ve(G)}\,,
\e{236}
\be
&&\{F,\{G, H\}\}(-1)^{\ve(F)\ve(H)}+
{\rm cycle}(F,G,H)=0\,.
\e{237}
Notice that \r{235} is satisfied by \r{230}
 by construction due to \r{223}, and
that \r{236} follows from \r{231}. The
 proof of \r{237} is given in \cite{Trip}.  The
nondegeneracy of $V^a$ should furthermore be
such that  the functions in $\cF$  span a
submanifold $\cL_1$ of $\cM$ with dimension
$2N$ (dim$\cM=6N$) and that the Poisson
bracket \r{230} is nondegenerate on $\cL_1$.
(In fact, the existence of a
symmetric $V^a$ {\em requires} these
nondegeneracy properties.) This generalizes the
conditions in \cite{Trip}.

\setcounter{equation}{0}
\section{New version of the $\Delta^a_\pm$-
operators and the origin  of $V^a$}
The generating  operators $\Delta_\pm^a$
introduced in \r{205} are composed of two
different objects, $\Delta^a$ and $V^a$,
 which have no natural relation between them
apart from the nilpotency conditions \r{208}
 of $\Delta_\pm^a$. In this section
we propose a unified expression for
$\Delta_\pm^a$ which is more invariant in the
sense that  \r{208} allows for
transformations of the form \r{238} for arbitrary
bosonic functions $\cH$.  In addition
 it allows us to demonstrate the existence of $V^a$
by deriving a geometrical explanation
for its origin. The resulting more explicit form
of $V^a$ is  also a covariant object
which incorporates the original expression in
Darboux coordinates obtained from
the Hamiltonian treatment in
\cite{BLT-2}-\cite{BM}.    Our
unified expression is \be
&&\Delta_\pm^a=\half(-1)^{\ve_A}
\L(\rho^{-1}\d_A\!\circ\!\rho\pm{i \over
\hbar}F_A\R)E^{ABa}\L(\d_B
\pm{i \over \hbar}F_B\R)
 \e{501}
where $F_A$
($\ve(F_A)=\ve_A$) is  an additional
connection to a trivial volume connection. $F_A$
transforms as a vector field. (One may
compare this expression with the
general ansatz for $\Delta$ in the
antisymplectic case given in
\cite{BT93}.) The operators \r{501}
are also hermitian with respect to the
scalar product \r{2071} provided $E^{ABa}$
transforms according to \r{2076}
and $F_A$ according to
\be
&&F_A=I^D_AF^*_D(-1)^{\ve_D},
\;\;\;F^*_A=F_DI^{*D}_A(-1)^{\ve_D}
\e{5011}
under the complex-conjugate condition \r{2072}.
Expanding \r{501} in
$i/\hbar$ we find  \be  &&\Delta_\pm^a=
\Delta^a\pm{i \over \hbar}(V^a+\half{\rm
div}V^a)+\L({i \over \hbar}\R)^2
\half(-1)^{\ve_A}F_AE^{ABa}F_B \e{502} where \be
&&V^a=V^{Aa}(-1)^{\ve_A}\d_A,\;\;\;V^{Aa}\equiv
E^{ABa}F_B=F_BE^{BAa}(-1)^{\ve_A+\ve_B}\,,  \e{503}
which, thus, is a special realization
 of the general $V^a$ of the previous section.
As mentioned above this special
 covariant $V^a$ also incorporates the original
expressions in special coordinates.
However,  due to the presence of the
$(i/\hbar)^2$ terms in \r{502} this
 $\Delta_\pm^a$  does not  coincide with the
$\Delta_\pm^a$ of the previous section.
Ignoring this fact for the moment we
consider the conditions  \be
&&
[\Delta_\pm^a,\Delta_\pm^b]=0\quad
\Leftrightarrow\quad
\Delta_\pm^{\{a}\Delta_\pm^{b\}}=0\,. \e{504}
To the zeroth order in $i/\hbar$
these conditions are identical to \r{209} or
equivalently \r{212} and \r{213},
and to the first order they yield \r{211} or
equivalently \r{215}-\r{217}.
Inserting \r{503} into \r{217} and making use of \r{212}
we find that \r{217} reduces to
\be
&&E^{AC\{a}\L(\d_CF_D-\d_DF_C(-1)^{\ve_C\ve_D}\R)
(-1)^{\ve_D}E^{DB|b\}}=0\,.
\e{505}
Conditions   \r{215}-\r{216} are however
complex relations beween $F_A$, $E^{ABa}$
and $\rho$ which are difficult
to solve in general coordinates.
 To the second and third order in $(i/\hbar)$
eq.\r{504} yields
\be
&\half[\Delta^{\{a},
(-1)^{\ve_A}F_AE^{AB|b\}}F_B]+[V^{\{a},
V^{b\}}]+\half[V^{\{a},{\rm div}\,V^{b\}}]=0\,,
\e{507}
\be
&[V^{\{a}, (-1)^{\ve_A}F_AE^{AB|b\}}F_B]=
(-1)^{\ve_A}(V^{\{a}F_AE^{AB|b\}}F_B)=0\,.
\e{508}
Identifying different powers of $\d_A$
in \r{507} it splits into the following
conditions
 \be
&(\Delta^{\{a}(-1)^{\ve_A}F_AE^{AB|b\}}F_B)+
(V^{\{a}{\rm div}\,V^{b\}})=0\,,
\e{509}
\be
\half E^{AB\{a}\d_B(-1)^{\ve_C}F_CE^{CD|b\}}
F_D+V^{\{a}V^{A|b\}}(-1)^{\ve_A}=0\,.
\e{510}
Remarkably enough  conditions \r{508}-\r{510}
turn out to be identically satisfied.
Eqs.\r{508} and \r{510} are satisfied  when
\r{505} and \r{212} are imposed, and
\r{509} follows from \r{216}. Thus,
eq.\r{505} are the only necessary conditions on
$F_A$ in order to satisfy \r{504} apart from
the involved relations   \r{215}-\r{216}
which might impose further restrictions.
This high degree of symmetry is also
reflected in the property that all
conditions coming from \r{504} are invariant under
the transformations  \be &F_A\;\ra\;F_A+\d_A\cH
\e{512}
for {\em any} bosonic function $\cH(\Gamma)$.
 Consider now the quantum master equation
\be
&\Delta^a_+\exp{\L({i\over\hbar}W\R)}=0\,,
\e{5121}
which may be written as
\be
&&\half(W,W)^a+(-1)^{\ve_A}(E^{ABa}F_B)\d_AW+
\half(-1)^{\ve_A}F_AE^{ABa}F_B=\nn\\
&&=i\hbar\Delta^aW+\half
i\hbar(-1)^{\ve_A}\rho^{-1}\d_A\rho E^{ABa}F_B\,.
\e{5122}
These equations are invariant under \r{512}
 provided we also shift the master action
$W$ by
\be
&W\mapsto W-\cH\,.
\e{5123}
For \r{5121} this is obvious from the form
\r{501} of $\Delta^a_+$. The classical part
of the master equation, \ie the left-hand
side of \r{5122}, does not agree with the
expression obtained from the Hamiltonian
treatment. This forces us to impose
 the auxiliary conditions
\be
&(-1)^{\ve_A}F_AE^{ABa}F_B=0\,.
\e{511}
They remove the last terms in \r{502} and
make our unified expression consistent
with
 the general conditions on $\Delta_\pm^a$
given in the previous
section.
This isotropy condition on $F_A$ is invariant
under the transformation \r{512}
provided the  bosonic function $\cH(\Gamma)$
 satisfies the classical master equation
\r{239} with $V^a$ given by \r{503}. Thus, the
invariance transformations \r{512} have
now been reduced to the   arbitrariness
\r{238} of $V^a$ in the
previous section.

One may now notice that the vorticity entering the parenthesis in
\r{505} can be
considered as   coefficients
of $d^{\{a}\omega_1^{b\}}$ where
$\omega^a_1$ is the one-form
\be
&&\omega^a_1\equiv F_Ad^a\Gamma^A\,.
\e{513}
We have
\be
&&d^{\{a}\omega_1^{b\}}=-\half\L(\d_AF_B-
\d_BF_A(-1)^{\ve_A\ve_B}\R)(-1)^{\ve_A}d^{\{b}\Ga^B\wedge
d^{a\}}\Ga^A\,. \e{5131}
Notice, however, that $\omega^a_1$ is not closed
since we cannot remove $E^{CAa/b}$
and $E^{BDb/a}$ in \r{505} (cf. \r{2131}).
The deviation from closeness is measured
by the tensor
\be
&&\omega^{AB}\equiv \half\epsilon_{ab}
E^{ACa}\L(\d_CF_D-
\d_DF_C(-1)^{\ve_C\ve_D}\R)(-1)^{\ve_D}E^{DBb}\,.
\e{514}
We know from our expressions of $V^a$
in terms of Darboux coordinates that
$\omega^{AB}$ is different from zero.
A condition which is invariant  under
\r{512} and which comply with these results are
\be
&& {\rm
rank}\:\omega^{AB}=2N,\;\;\;({\rm dim} \cM=6N)\,.
 \e{515}
This is just a nondegeneracy condition
 on $F_A$. It means that $F_A$ must be nontrivial
($F_A\neq\d_A\cH$) and that $V^a$ cannot
be transformed away in the quantum master
equation \r{2412}. It is related to our
 condition of a nondegenerate
Poisson bracket given in \cite{Trip} and
in the previous section.  Indeed if we specify
the submanifold $\cL_1$ of $\cM$
 appropriate for the Poisson bracket \r{230}
 by  $F_A=0$  we have
\be
&&\L.\{\Gamma^A, \Gamma^B\}\R|_{\cL_1}=\omega^{AB}\,.
\e{516}
Another invariant condition which comply
with the results in Darboux coordinates is
\be
&& {\rm
rank}\:F_{AB}=4N,\;\;\;F_{AB}\equiv
\d_AF_B-\d_BF_A(-1)^{\ve_A\ve_B}\,.
 \e{517}
 A still stronger invariant condition
also true in special coordinates is
\be
&&(-1)^{\ve_B}F_{AB}E^{BCa}F_{CD}=0\,,
\e{518}
which has the structure of the isotropy
 condition \r{511}. It follows immediately from
\r{518} that
\be
&&\omega^{AB}F_{BC}=0\,.
\e{519}
 We could  require for
$\omega^{AB}$ and $F_{AB}$ to form  complete
 basis for each others nullvectors so that
\be
&&{\rm
rank}\:\omega^{AB}+{\rm
rank}\:F_{AB}=6N\,,
\e{520}
which is  consistent with \r{515} and \r{517}.

If the condition \r{518} is accepted
we may introduce a new Poisson bracket defined on the whole
triplectic manifold $\cM$ by
\be
&&\{F,G\}\equiv F{\stackrel{\lea}
{\partial_A}}\omega^{AB}\d_BG
\e{521}
for {\em arbitrary} functions $F(\Ga)$ and $G(\Ga)$.
It satisfies the antisymmetry \r{235}
and the Leibniz rule \r{236}. The Jacobi
identities \r{237} follow from the properties
\be
&&\omega^{AD}\d_D\omega^{BC}
(-1)^{\ve_A\ve_C}+{\rm cycle}(A,B,C)=0\,,
\e{522}
which are straight-forward to prove.
First one has to use the cyclic
identities
\be
&&\d_AF_{BC}(-1)^{\ve_A\ve_C}+
{\rm cycle}(A,B,C)\equiv0,
\e{523}
then conditions \r{505} and \r{518} to
 remove the derivatives on $F_{AB}$. Finally one
has to use \r{212} to reproduce \r{522}.
The Poisson bracket \r{521} is degenerate on
$\cM$. It is nondegenerate only on the
submanifold $\cL_1$ defined by $F_A=0$ which
is of dimension $2N$. Notice that \r{521}
is equivalent to \r{230} only on $\cL_1$.
Notice also that the condition \r{232}
reduces to \r{505} on $\cL_1$. In distinction
to the Poisson bracket \r{230} the new
bracket \r{521} is invariant under the gradient
shifts \r{512} for {\em any} bosonic
generator $\cH$. The properties of the submanifold
$\cL_1$ are in general rather complicated as it may
be seen from the generalized involution
relations   for $F_A$  which follow from
\r{511} and \r{518}, \ie
 \be
&&(F_A, F_B)^a+(-1)^{\ve_C}F_{AC}(\Gamma^C,
F_B)^a-\nn\\&&-(-1)^{\ve_C}F_{BC}(\Gamma^C,
F_A)^a(-1)^{(\ve_A+1)(\ve_B+1)}=\cU^{Ca}_{AB}F_C\,.
\e{524}
Notice that  these relations   are only
invariant under the
same restricted shifts \r{512} under which
the isotropy conditions \r{511} are
invariant.

The basic reason why we have  $V^a$ operators
within  triplectic quantization and
not corresponding $V$ operators within the
antisymplectic quantization is the fact that
\r{505} allows for nontrivial solutions. In
antisymplectic quantization the corresponding
condition to \r{505} may be transformed
to $\d_CF_D-\d_DF_C(-1)^{\ve_C\ve_D}=0$ with
 only trivial solutions which may be
transformed away in the master equation.

 \setcounter{equation}{0} \section{Gauge fixing
in general triplectic
quantization}

The path integral in general
triplectic Lagrangian quantization is proposed
to be \cite{Trip}
\be
&&Z=\int\exp\biggl\{{i\over\hbar}\Bigl[W +
 X\Bigr] \biggr\}
\rho(\Gamma)[d\Gamma][d\lambda]\,,
\e{301}
where $W(\Gamma; \hbar)$ is the quantum
master action which satisfies \r{2412} and where
 $X(\Gamma, \la; \hbar)$ is a gauge
fixing action which depends on the parametric
 variables $\la^\al$, $\al=1,\ldots,N$,
which are generalized Lagrange multipliers for
 hypergauge conditions.  In \cite{Trip}
it was shown that provided $X$ satisfies the
"weak" quantum master equation
\be
\Bigl(\Delta^a-\frac{i}{\hbar}\cV^a-
\frac{i}{\hbar}(-1)^{\ve_{\al}}R^{\al
a}\d_{\al}+ (-1)^{\ve_{\al}}\d_{\al}R^{\al a}
\Bigr)\exp{\L(\frac{i}{\hbar}X\R)}=0\,,
\e{302}
or equivalently,
\be \half(X,
X)^a-i\hbar\Delta^a X-V^aX+\half i\hbar\, {\rm div
}\,V^a-X{\!\stackrel{\lea}{\d}}_{\al}R^{\al a}
+ i\hbar R^{\al
a}{\!\stackrel{\lea}{\d}}_{\al}= 0\,, \e{303}
where  $\d_{\al}\equiv\d/\d\la^{\al}$ and
$\ve_{\al}\equiv\ve(\la^{\al})$,
then \r{301} is invariant under the general
 canonical transformation
\be
&&\delta\Gamma^A={}(\Gamma^A, -W+X)^a\mu_a -
 2V^{A a}\mu_a\,,\;\;\;
\delta\la^{\al}=-2R^{\al a}\mu_a\,,
\e{304}
 where $\mu_a$ are
two fermionic constants.
The consistency conditions for \r{302} are
\be{\d\over{\d\la^\al}}
\L\{\L(i\hbar(\Delta^{\{a}R^{\al b\}})+
V^{\{a}R^{\al b\}}-(X,
R^{\al\{ a})^{b\}}+(-1)^{\ve_\beta}R^{\beta
\{a}\d_\beta R^{\al b\}}\R)
e^{\frac{i}{\hbar}X}\R\}=0\,,
\e{305}
which are obtained by applying the operator
\be
\Delta^b-\frac{i}{\hbar}V^b+
\frac{i}{\hbar}(X,\,\cdot\,)^b-
\frac{i}{\hbar}(-1)^{\ve_\al}R^{\al b}\d_\al
\e{306}
to the left-hand side of eq.\r{302}
and symmetrizing in $a$ and $b$.
As shown in \cite{Trip} \r{305} may be
solved within an
extended formalism which is obtained as follows:
Introduce a linear space $\Lambda$ spanned
by $(\la^\al, \la^*_{\al a},
\bar\la_\al, \eta^{\al a})$, with
$\epsilon(\la^\al)=\epsilon(\bar\la_\al)=
\epsilon_\al$, $\epsilon(\la^*_{\al
a})=\epsilon(\eta^{\al a})=\epsilon_\al+1$,
and define an extended triplectic
manifold $\tilde\cM=\cM\times\Lambda$\@.
On $\tilde\cM$, one may then introduce
the operators
\be
\Delta^a_{\rm ext}=
\Delta^a+(-1)^{\ve_\al}{\d\over{\d\la^\al}}
{\d\over{\d\la^*_{\al a}}} +
(-1)^{\ve_\al+1}\epsilon^{ab}
{\d\over{\d\bar\la_\al}}{\d\over{\d\eta^{\al b}}}
\e{307}
 and the corresponding
antibrackets
\be
 &&(F,G)^a_{\rm ext}= (F,G)^a +\nn\\
&&+\Bigl({F}{\stackrel{\lea}{\d}
\over{d\la^\al}}{\d\over{\d\la^*_{\al a}}}{G} +
\epsilon^{ab}{F}
{\stackrel{\lea}{\d}\over{\d\eta^{\al b}}}
{\d\over{\d\bar\la_\al}}{G}
-(-1)^{(\ve(F)+1)(\ve(G)+1)}
(F\leftrightarrow G)\Bigr)\,. \e{308}
The vector fields $V^a$ may then be
 extended to $\tilde\cM$ \eg as follows (which
generalizes \cite{Trip})
\be
 \caV^a \equiv V^a -
(1-\beta) \epsilon^{ab}\la^*_{\al b}
{\d\over{\d\bar\la_\al}} +\beta
(-1)^{\ve_\al}\eta^{\al a}{\d\over{\d\la^\al}}\,,
\e{309}
which satisfy the necessary conditions
$[\hat\caV^{\{a},\,\Delta^{b\}}_{\rm ext}]
=0$ etc.\ in accordance with the conditions
given in the previous section for any
 real constant $\beta$.

Now we can introduce an extended quantum master equation
\be
 \L(\Delta^a_{\rm
ext}-\frac{i}{\hbar}\hat\caV^a\R)\exp{
\L(\frac{i}{\hbar}\cX\R)}=0,\;\;\;
\hat\caV^a\equiv\caV^a+\half
{\rm div}\,\caV^a\,,
 \e{310}
 or equivalently,
\be
\half(\cX,\,\cX)^a_{\rm ext}-\caV^a\cX-
i\hbar\Delta^a_{\rm
ext}\cX+\half i\hbar\,{\rm div}\,
\caV^a=0\,.
\e{311}
 The "weak" master equation \r{302} follows then
from \r{310} or \r{311} once we
 take $\cX$ to be
\be
 \cX(\Gamma, \la, \la^*, \bar\la,
\eta) = \tX(\Gamma, \la, \la^*, \bar\la) +
\beta\la^*_{\al a}\eta^{\al a}
\e{312}
 and  expand $\tX$ as follows
\be
 &&\tX(\Gamma, \la, \la^*, \bar\la) =
 X(\Gamma,
\la) - \la^*_{\al a}R^{\al a}(\Gamma, \la) -
 \bar\la_\al\bar R^\al(\Gamma, \la) +
\nn\\
&&+\half\la^*_{\al a}\la^*_{\beta b} F^{\al a;\beta b} +
\half\bar\la_\al\bar\la_\beta
\bar F^{\al\beta} + \bar\la_\beta\la^*_{\al a}E^{\al
a;\beta} + {\rm higher\ orders\ in\ }\la^*,\bar\la\,.
\e{313}
Eq.\r{310} implies then \eg to
 the first order in $\la^*$:
\be
&&i\hbar\Delta^a R^{\al b} + V^aR^{\al b} +
(-1)^{\ve_\beta}R^{\beta a}\d_\beta R^{\al
b} - (X,\,R^{\al b})^a + (-1)^{\ve_\al}
\epsilon^{ab}\bar R^\al =\nn\\
&&=(-1)^{\ve_\beta}(i\hbar\d_\beta
F^{\al b;\beta a} - \d_\beta X F^{\al b;\beta a})\,,
\e{315}
which solves the consistency conditions \r{305}
upon symmetrization.

The integral \r{301} can now be reformulated on
the extended triplectic manifold using
the extended master action $\cX$ as follows
\cite{Trip}
\be
Z
&=&\int\exp\biggl\{{i\over\hbar}
\Bigl[W + \tilde{X} + \la^*_{\al a}\eta^{\al a} +
\bar\la_\al\xi^\al \Bigr] \biggr\}\rho(\Gamma)
\underbrace{[d\Gamma][d\lambda][d\lambda^*]
[d\bar\lambda][d\eta]}_{[d\tilde\Gamma]}[d\xi]\nn\\
&=&\int\exp\biggl\{{i\over\hbar} \Bigl[\cW + \cX \Bigr]
\biggr\}\rho(\tilde\Gamma)[d\tilde\Gamma][d\xi]
\e{316}
where $\xi^\al\equiv\la^{(1)\al}$ are
 Lagrange multipliers of the {\it next-
level\/} theory, and where
\be
\cW=W+(1-\beta)\la^*_{\al a}\eta^{\al a} +
 \bar\la_\al\xi^\al
\e{317}
is a{ solution\/} on
$\tilde\cM$ to the strong master equation:
\be
&&\L(\Delta^a_{\rm
ext}+\frac{i}{\hbar}\hat\caV^a\R)\exp{
\L(\frac{i}{\hbar}\cW\R)}=0\,,
\e{3171}
or equivalently,
\be
\half(\cW,\,\cW)^a_{\rm ext} + \caV^a\cW -
i\hbar\Delta^a_{\rm ext}\cW-\half
i\hbar\,{\rm div}\,\caV^a=0\,.
 \e{318}
Notice that at this level $\cW$ has
 become a gauge fixing action (depends on
$\xi^\al$) and $\cX$ a master action.

The path integral \r{316} is
 invariant under the canonical transformation
\be
&&\delta\tGamma^I={}(\tGamma^I, -\cW+
\cX)_{\rm ext}^a\,\mu_a - 2\caV^{I
a}\mu_a\,,\;\;\; \delta\la^{(1)\al}
\equiv\delta\xi^\al=0\nn\\
&&I=1,\ldots,12N,\;\;\;\al=1,\ldots,N
\e{320}
where $\mu^a$ again are two fermionic constants.
One may notice that the natural arbitrariness
in $\caV^a$ as described by the
transformations
\be
&&\caV'^a=\caV^a+(\cH, \cdot)^a_{\rm ext},
\;\;\;\half(\cH, \cH)_{\rm
ext}^a+\caV^a\cH=0\,, \e{321}
which generalize formulas \r{238} and \r{239},
 and where  $\cH(\tilde\Gamma)$ here is a
bosonic function on $\tilde\cM$, imply that
the master equations \r{310} and \r{3171}
are invariant under $\caV^a\mapsto\caV'^a$
 provided $\cW$ and $\cX$ at the same time
are shifted according to  $\cW\mapsto\cW'=
\cW-\cH$, $\cX\mapsto\cX'=\cX+\cH$. In
\r{316} we get then $\cW+\cX\mapsto\cW'+
\cH'=\cW+\cX$ and the invariance transformations
\r{320} become exactly the same expressions
with $\cW$, $\cX$ and $\caV^a$ replaced by
$\cW'$, $\cX'$ and $\caV'^a$. In fact,
the arbitrariness in $\caV^a$ given in \r{309}
as represented by the arbitrary constant
$\beta$ is of the general form \r{321} since
\be
&&\caV^a \equiv V^a  -\epsilon^{ab}\la^*_{\al b}
{\d\over{\d\bar\la_\al}}+(\beta\la^*_{\al b}\pi^{\al
b},\,\cdot\,)^a \,.\e{322}

The solutions of \r{310} and \r{3171} are  mapped
  on new
solutions, $\cX\mapsto\cX'$ and
 $\cW\mapsto\cW'$, according to the formulas
\cite{BLT-4}
\be
 &&\exp{\L(\frac{i}{\hbar}\cX'\R)}=\exp\L(
\epsilon_{ab}\frac{\hbar}{i}[
\Delta^a_{\rm ext}-\frac{i}{\hbar}\hat\caV^a,\,
[\Delta^b_{\rm ext}-\frac{i}{\hbar}
\hat\caV^b,\Phi]\,]\R)
\exp{\L(\frac{i}{\hbar}\cX\R)}\,,
\e{323}
\be
 &&\exp{\L(\frac{i}{\hbar}\cW'\R)}=\exp\L(
\epsilon_{ab}\frac{\hbar}{i}[
\Delta^a_{\rm ext}+\frac{i}{\hbar}
\hat\caV^a,\,
[\Delta^b_{\rm ext}+\frac{i}{\hbar}
\hat\caV^b,\Psi]\,]\R)
\exp{\L(\frac{i}{\hbar}\cW\R)}\,,
\e{324}
 where $\Phi$ and $\Psi$ are arbitrary
{\em bosonic} functions or operators.
In \cite{Trip} a proof was given that the path
integral \r{316} is independent of
the natural arbirariness of the solutions of
\r{310} given in \r{323}. This proof was
 based on the transformation \r{320}. The path
integral is also independent of the
 arbirariness \r{324} respecting the boundary
conditions \r{2413}.

\setcounter{equation}{0}
\section{Second class hyperconstraints}
As shown in \cite{Trip} if we set the
 classical limit of the gauge fixing action $X$
in \r{301} to be linear in $\la^\al$, \ie
\be
&&X(\Gamma,\la;0)=G_\al(\Gamma)\la^\al+Z(\Gamma)\,,
\e{401}
then the "weak" master equation \r{302} requires
\be
&&(G_\al, G_\beta)^a=G_\ga U_{\al\beta}^{\ga a}\,,
\e{402}
\be
&&(Z, G_\al)^a-V^aG_\al= G_\beta U_{\al}^{\beta a}\,,
\e{4021}
\be
&&\half(Z, Z)^a-V^aZ=\half G_\ga U^{\ga a}\,,
\e{4022}
where $U_{\al\beta}^{\ga a}$, $U_{\al}^{\beta a}$
and $ U^{\ga a}$ are coefficients
for different powers of $\la^\al$ in
$R^{\ga a}(\Gamma,\la;0)$ (see \cite{Trip}).
One may notice that $Z$ may be
transformed away in $X$ by a transformation
of the form \r{321} after which
\r{4022} is eliminated. $G_\al$ are here
hypergauge generators which may be viewed as
first class hyperconstraints due to \r{402}.
By means of the nondegeneracy concept
introduced for $E^{ABa}$ in \r{202}-\r{204}
 we may also introduce second class
hyperconstraints within the triplectic
framework. We have then the following
generalization of the corresponding
antisymplectic treatment of \cite{BT93}: Call
$\Theta^\al(\Gamma)$, $\al=1,\ldots,6K$,
 second class constraints
if there exists a $Y^{c}_{ab\beta\ga}$
satisfying \be &&E^{\al\beta a}
Y^{c}_{ab\beta\ga}=\del^\ga_\al\del^c_b,
\;\;\;E^{\al\beta a}\equiv(\Theta^\al,
\Theta^\beta)^a\,, \e{403} and  \be
&&\ve(Y^{c}_{ab\beta\ga})=\ve(\Theta^\beta)+
\ve(\Theta^\ga)+1,\;\;\;Y^{a}_{bc\beta\ga}
=-Y^{a}_{cb\ga\beta}(-1)^{
\ve(\Theta^\beta)\ve(\Theta^\ga)}\,.
\e{404}
We may then define the triplectic
counterpart of the Dirac bracket by
\be
&&(A, B)^a_{(D)}\equiv(A, B)^a-(A,
\Theta^\beta)^bY^{a}_{bc\beta\ga}(\Theta^\ga,B)^c\,.
\e{405}
These generalized Dirac brackets
 satisfy all the required
antibracket properties \r{222}-\r{227} and
\be
&&(A, \Theta^\beta)^a_{(D)}\equiv 0\,.
\e{4051}

Notice now that
\be
&&E^{ABa}_{(D)}\equiv(\Gamma^A, \Gamma^B)^a_{(D)}
\e{406}
is an example of a degenerate tensor.
Still we may introduce nilpotent differential
operators
 \be
&&\Delta_{\pm(D)}^a=\Delta_{(D)}^a\pm
{i\over\hbar}(V^a_{(D)}+\half{\rm
div}V^a_{(D)}),\nn\\
&&\Delta^a_{(D)}=\half(-1)^{\ve_A}
\rho_{(D)}^{-1}\d_A\!\circ\!\rho_{(D)} E^{A B
a}_{(D)}\d_B\,,\nn\\  &&
V^a_{(D)}=(-1)^{\ve_A}V^{Aa}_{(D)}\d_A,\;\;\;{\rm
div}V^a_{(D)}\equiv\rho_{(D)}^{-1}
\L(\d_A\rho_{(D)} V^{Aa}_{(D)}\R)(-1)^{\ve_A}\,,
\e{407}
which satisfy the required nilpotency
conditions corresponding to \r{208},
and which, furthermore,  satisfy
\be
&&\Delta_{(D)}^a\Theta^\al=0,\;
\;V^a_{(D)}\Theta^\al=0\;
\;\Rightarrow\;\;\Delta_{\pm(D)}^a\Theta^\al=0\,.
\e{408}
The last properties follow from \r{4051}
 if we make use of the natural representation
\be
&&V^a_{(D)}=(-1)^{\ve_B}F_BE^{BAa}_{(D)}\d_A\,,
\e{409}
in accordance with \r{503}.  In this
latter case $\Delta_{\pm(D)}^a$
may also be written in the unified form \r{501}
 with $E^{ABa}$ replaced
by $E^{ABa}_{(D)}$.

The second
 class constraints, $\Theta^\al=0$, should
eliminate $6K$ degrees of freedom from
the original triplectic manifold $\cM$ (dim
$\cM=6N$) leaving a nondegenerate
triplectic manifold $\cM_{(D)}$ of dimension
$6(N-K)$. Corresponding to \r{301}
we have then the gauge fixed path integral
expression \be &&Z=\int\exp
\biggl\{{i\over\hbar}\Bigl[W_{(D)} +
X_{(D)}\Bigr] \biggr\}
d\mu_{(D)}(\Gamma)[d\lambda]\,,
\e{410}
where $W_{(D)}$ and $X_{(D)}$ satisfy the 
master eqs.\r{2412} and \r{303} with
 all operators replaced by their Dirac
counterparts, and where the volume
 measure is given by \be
&&d\mu_{(D)}(\Gamma)\equiv\rho_{(D)}(
\Gamma)\prod_\al\del(\Theta^\al)[d\Gamma]\,.
 \e{411}
The path integral \r{410} is then
invariant under the general
 canonical transformation
\be
&&\delta\Gamma^A={}(\Gamma^A,
-W_{(D)}+X_{(D)})^a_{(D)}\mu_a -
 2V^{A a}_{(D)}\mu_a\,,\;\;\;
\delta\la^{\al}=-2R^{\al a}\mu_a\,,
\e{412}
 where $\mu_a$ are
two fermionic constants.
  \newpage
{\bf Acknowledgements}\\

I.B. is thankful to Lars Brink for
warm hospitality at the Institute of Theoretical
Physics, Chalmers and G\"{o}teborg
University. The work of I.B. is partially
supported by Grant \#M2I300 from
the International Science Foundation and the
Government of Russian Federation,
by Human Capital and Mobility Program of the
European Community under Projects
INTAS 93-0633, 93-2058, by NATO Linkage Grant
\#931717, and by Research Grant from
the British Research Council for Astronomy and Particle
Physics. I.B. and R.M. are thankful
to the Royal Swedish Academy of Sciences for
financial support.\\ \\

\setcounter{section}{1}
\setcounter{equation}{0}
\renewcommand{\thesection}{\Alph{section}}

    \noindent
    {\Large{\bf{Appendix}}}\\ \\
 {\bf Some basic objects as represented in terms of the
Darboux coordinates}\\

Let us consider the special Darboux coordinates
 of refs.\cite{BM} and \cite{Trip},
\ie  \be
&&\Gamma^A=(\phi^\al, \phi^*_{\al a}),
\;\;\al=1,\ldots,2N,\;\;\;
\ve(\phi^\al)=\ve_\al,\;\;
\ve(\phi^*_{\al a})=\ve_\al+1\,,
\e{a1}
where $\phi^\al$ are field variables and
$\phi^*_{\al a}$ antifields.
In terms of these coordinates we have
\be
&&\Delta^a=(-1)^{\ve_\al}\frac{\d}{\d
\phi^\al}\frac{\d}{\d\phi^*_{\al a}}\,,
\e{a2}
\be
 &&(F,G)^a=F\frac{\stackrel{\lea}{\d}}{\d
\phi^\al} \frac{\d}{\d\phi^*_{\al a}}G -
 F\frac{\stackrel{\lea}{\d}}{\d
\phi^*_{\al a}} \frac{\d}{\d\phi^\al}G
\,,
\e{a3}
and
\be
&&V^a=\epsilon^{ab}\phi^*_{\al b}\,
\kappa^{\al\beta} \frac{\d}{\d\phi^\beta}\,,
\e{a4}
where $\kappa^{\al\beta}$ is an off-diagonal
constant matrix.  The last expression includes
the $V^a$ operators given in
refs.\cite{BLT-2}-\cite{BM}.

The basic antitriplectic metric
$E^{ABa}$ is given by
\be
&&E^{ABc}=\L(\bea{cc} E^{\al\beta c}
&{{E^{\al}}_{ \beta b}}^ c\\
{E_{\al a}}^ {\beta\: c}&{E_{\al a\:
 \beta b}}^{ c}\ea\R)=\L(\bea{cc}
0&\del^\al_\beta\del^c_b\\
-\del_\al^\beta\del^c_a&0\ea\R)\,,
\e{a7}
and the inverses $Y^c_{dfAB}$ are in turn
\be
&&Y^c_{dfAB}=\L(\bea{cc} Y^c_{df\al\beta}
&{Y^c_{df\al}}^{\beta b}\\
{{Y^c_{df}}^{\al a}}_{\beta}&
{Y^c_{df}}^{\al a \:\beta b}\ea\R)=\L(\bea{cc}
0&-\del^\beta_\al \del^b_d \del^c_f\\
\del^\al_\beta\del^a_f\del^c_d&
{Y^c_{df}}^{\al a \:\beta b}\ea\R)\,,
\e{a8}
where ${Y^c_{df}}^{\al a \:\beta b}$ is not
uniquely determined. The conditions
 \r{202} and \r{203} only require
\be
&&{Y^c_{af}}^{\al a \:\beta
b}=0\,.\e{a9}
The two-form $\omega_2^a$
defined by \r{2133} is then ($D^a=d^a$ in Darboux coordinates)
\be
&\omega_2^a=& (-1)^{\ve_\beta}d^a\phi^*_{\beta b}\wedge
d^b\phi^\beta
-\half{Y^a_{df}}^{\beta b \:\ga
c}(-1)^{\ve_\ga}\,d^f\phi^*_{\ga c}\wedge d^d\phi^*_{\beta
b}-\nn\\ &&-{F_{b}}^{\ga c}d^ad^b\phi^*_{\ga c}
-F_{b\beta}d^ad^b\phi^\beta\,,
 \e{a10}
where ${F_{b}}^{\ga c}$ and $F_{b\beta}$ are also undetermined. 
This arbitrariness may be fixed if we  assume that $\omega_2^a$ is
exact for Darboux
 coordinates. From \r{2134}, \r{2135} we have then more precisely
\be
&&\omega_2^a=d^a\omega=(-1)^{\ve_\beta}d^a\phi^*_{\beta b}\wedge
d^b\phi^\beta-\phi^*_{\beta b}d^ad^b\phi^\beta,
\;\;\;\omega=\phi^*_{\al a}d^a\phi^\al(-1)^{\ve_\al}\,.
\e{a100}

If we consider the unified expression \r{501}
 of section 3 we may easily derive the
general nontrivial form of $V^a$ in terms
of the Darboux coordinates. We start then
from the natural representation \r{503}
of $V^a$, \ie
\be
&&V^a=(-1)^{\ve_B}F_BE^{BAa}\d_A=
(-1)^{\ve_\al}F_\al
{\d \over \d\phi^*_{\al a}}+(-1)^{\ve_\al}
F^{\al a}{\d \over \d\phi^\al}\,,
\e{a11}
where we in the last equality have made
 use of  \r{a1} and \r{a7}.
Inserting  $F_A$ into the condition
\r{505} using \r{a7} we find that the
nontrivial part of $F_A$ ($\neq\d_A\cH$)
is contained in the antifield components and
that these components only depend on the
 antifields, \ie $F^{\al a}(\phi^*_{\beta
b})$. A general linear ansatz for
$F^{\al a}$ satisfying \r{505} is then
\be
&&F^{\al a}=\epsilon^{ab}\phi^*_{\beta b}
\kappa^{\beta\al}(-1)^{\ve_\al}\,,
\e{a12}
which when together with $F_\al=0$ are
inserted into \r{a11} exactly reproduces
\r{a4}. Different matrices
$\kappa^{\al\beta}$  are related by
the gradient shifts \r{512}  apart
from the absolute
normalization. We have
\be
&&F^{\al a}\,\mapsto\,F^{\al a}+
{\d\cH\over\d\phi^*_{\al a}}\,,
\e{a16}
where the most general form of $\cH$ is
\be
&&\cH=\half\epsilon^{ab}\phi^*_{\al a}
\sigma^{\al\beta}\phi^*_{\beta b}\,,
\e{a17}
where $\sigma^{\al\beta}$ is a constant
 symmetric matrix, \ie
$\sigma^{\al\beta}=
\sigma^{\beta\al}(-1)^{\ve_\al\ve_\beta}$.
 The shift \r{a16} is then
equivalent to $\kappa^{\al\beta}\,
\mapsto\,\kappa^{\al\beta}+\sigma^{\al\beta}$.
This generalizes the gradient shifs considered for
$\caV^a$ in \r{309}, \r{322}.

The next condition involves
 $\omega^{AB}$ defined by
\r{514}. In terms of the Darboux coordinates
\r{a1} $\omega^{AB}$ becomes
\be
&&\omega^{AB}=\L(\bea{cc} \omega^{\al\beta}
&{\omega^{\al}}_{ \beta b}\\
{\omega_{\al a}}^{ \beta}&\omega_{\al a
\:\beta b}\ea\R)=\L(\bea{cc} \omega^{\al\beta}
&0\\
0&0\ea\R)\,,
\e{a5}
where \be
&\omega^{\al\beta}&=\half\epsilon_{ab}\L({\d
F^{\beta b}\over \d \phi^*_{\al a}}-{\d
F^{\al a}\over \d \phi^*_{\beta
b}}(-1)^{(\ve_\al+1)(\ve_\beta+1)}
\R)(-1)^{\ve_\beta}=\nn\\
&&=\kappa^{\al\beta}-\kappa^{\beta\al}
(-1)^{\ve_\al\ve_\beta}\,.
 \e{a13}  The maximal rank of
$\omega^{\al\beta}$ is
obviously $2N$ since the number of field
 components are $2N$ according to \r{a1}. This
is also the rank that follows from
the $V^a$ of  refs.\cite{BLT-2}-\cite{BM}.

The
Poisson bracket \r{521} is defined in terms
 of $\omega^{AB}$ and due to \r{a5} it is
here given by  \be
&&\{F, G\}=F\stackrel{\lea}{\d}_\al
\omega^{\al\beta}\d_\beta G\,,
\e{a6}
which is nondegenerate on the $2N$
dimensional field manifold $\cL_1=\{\phi^\al\}$.
One may notice that $\kappa^{\al\beta}$ in the symmetric
$V^a$ of
\cite{BM} and \cite{Trip} satisfies
$\kappa^{\al\beta}=-\kappa^{\beta\al}
(-1)^{\ve_\al\ve_\beta}$ which implies
$\omega^{\al\beta}=2\kappa^{\al\beta}$ and
that this antisymmetric $\kappa^{\al\beta}$
cannot be modified by the gradient shift
\r{a16} since  $\sigma^{\al\beta}$
in \r{a17} is a symmetric matrix.

Finally we give  the one-form
\r{513} in the Darboux coordinates \r{a1}.
 From \r{a12} we find
\be
&&\omega_1^a=\epsilon^{cb}\phi^*_{\beta
b}\kappa^{\beta\al}(-1)^{\ve_\al}
d^a\phi^*_{\al c}+d^a\cH\,,
\e{a14}
where $\cH$ satisfies the classical master equation,
and \r{5131} reduces to
\be
&&d^{\{a}\omega_1^{b\}}=-\half\epsilon^{cd}
\omega^{\al\beta}(-1)^{\ve_\al+\ve_\beta}d^{\{b}
\phi^*_{\beta d}\wedge d^{a\}}\phi^*_{\al
c}\,.
\e{a15}\\
\\

    \end{document}